\documentclass[twocolumn]{aastex631}

\received{August 31, 2023}
\submitjournal{RN AAS}

\defcitealias{Rybak2020a}{R20}

\begin{document}

\title{Bright beacons? ALMA non-detection of a supposedly bright [OI] 63-$\mu$m line in a redshift-6 dusty galaxy}

\shorttitle{Non-detection of the [OI] 63-$\mu$line in G09.83808}
\shortauthors{M. Rybak et al.:}

\author[0000-0002-1383-0746]{Matus Rybak}
\email{m.rybak@tudelft.nl}
\affiliation{Faculty of Electrical Engineering, Mathematics and Computer Science, Delft University of Technology, Mekelweg 4, 2628 CD Delft,
The Netherlands}
\affiliation{Leiden Observatory, Leiden University, P.O. Box 9513, 2300 RA Leiden, The Netherlands}
\affiliation{SRON - Netherlands Institute for Space Research, Niels Bohrweg 4, 2333 CA Leiden,
The Netherlands}

\author{L. Lemsom}
\affiliation{Faculty of Applied Sciences, Delft University of Technology, Lorentzweg 1, 2628 CJ Delft, The Netherlands}
\affiliation{Faculty of Electrical Engineering, Mathematics and Computer Science, Delft University of Technology, Mekelweg 4, 2628 CD Delft,
The Netherlands}

\author{A. Lundgren}
\affiliation{European Southern Observatory, Karl Schwarzschild Stra\ss e 2, 85748 Garching, Germany}
\affiliation{Aix Marseille Université\'e, CNRS, LAM, Marseille, F-13388, France}

\author{J. Zavala}
\affiliation{National Astronomical Observatory of Japan, 2-21-1 Osawa, Mitaka, Tokyo 181-8588, Japan}

\author{J. A. Hodge}
\affiliation{Leiden Observatory, Leiden University, P.O. Box 9513, 2300 RA Leiden, The Netherlands}

\author{C. de Breuck}
\affiliation{European Southern Observatory, Karl Schwarzschild Stra\ss e 2, 85748 Garching, Germany}
\author{C. M. Casey}
\affiliation{Department of Astronomy, The University of Texas at Austin, 2515 Speedway Boulevard Stop C1400, Austin, TX 78712, USA}
\author{R. Decarli}
\affiliation{INAF Observatory of Astrophysics and Space Science of Bologna, via Gobetti 93/3, 40129, Italy}

\author{K. Torstensson}
\affiliation{Joint ALMA Observatory, Alonso de Córdova 3107, Vitacura 763-0355, Santiago de Chile, Chile}
\author{J. L. Wardlow}
\affiliation{Physics Department, Lancaster University, Bailrigg, Lancaster, LA1 4YB, UK}
\author{P. P. van der Werf}
\affiliation{Leiden Observatory, Leiden University, P.O. Box 9513, 2300 RA Leiden, The Netherlands}

\begin{abstract}
We report a non-detection of the [\ion{O}{1}] 63-$\mu$m emission line from the $z=6.03$ galaxy G09.83808 using ALMA Band~9 observations, refuting the previously claimed detection with APEX by \citep{Rybak2020a}; the new upper limit on the [\ion{O}{1}] 63-$\mu$m flux is almost 20-times lower. [\ion{O}{1}] 63-$\mu$m line could be a powerful tracer of neutral gas in the Epoch of Reionisation: yet our null result shows that detecting [\ion{O}{1}] 63-$\mu$m from $z\geq6$ galaxies is more challenging than previously hypothesised.

\end{abstract}


\section{Introduction}
\label{sec:introduction}

The [\ion{O}{1}] 63-$\mu$m line (henceforth [\ion{O}{1}]$_{63}$, rest-frame frequency 4744.78~GHz) is often the brightest emission line of the neutral gas in star-forming galaxies, outshining the [\ion{C}{2}]$_{158}$ line by a factor of a few (e.g. \citealt{Farrah2013}. With the rapidly mounting detections of galaxies at $z\geq6$, [\ion{O}{1}]$_{63}$ presents a potentially powerful ``bright beacon'' for measuring spectroscopic redshifts and studying gas content of galaxies in the Epoch of Reionisation.
Although [\ion{O}{1}]$_{63}$ has been detected in a number of $z=1-3$ galaxies (e.g., \citealt{Sturm2010,Coppin2012, Brisbin2015, Zhang2018b, Wagg2020}), to date, there has been just a single claimed [\ion{O}{1}]$_\mathrm{63}$ detection at $z\geq3$: in \citet[henceforth R20]{Rybak2020a}, we reported a detection of the [\ion{O}{1}]$_{63}$ line in G09.83808, a strongly lensed, dusty star-forming galaxy at redshift $z=6.02$ \citep{Zavala2018}, using the SEPIA660 spectrometer \citep{Baryshev2015, Belitsky2018} on the APEX telescope.

The SEPIA660 data showed a significant ($\approx$5.2$\sigma$) signal at 675.45~GHz ($z_\mathrm{[OI]}=6.0246$, consistent with the systemic redshift $z=6.0244\pm0.0003$, \citealt{Tsujita2022}). The detection was reproduced by two team members using several approaches. The measured line flux was 22$\pm$5 Jy km s$^{-1}$ which corresponds to a sky-plane (lensing-uncorrected) luminosity of $L_\mathrm{[OI]63}=(5.4\pm1.2)\times10^{10}$ $L_\odot$; almost four times brighter than the [\ion{C}{2}] 158-$\mu$m line \citep{Zavala2018}.

\section{Observations and data reduction}
\label{sec:observations}

To exploit the seemingly bright [\ion{O}{1}]$_\mathrm{63}$ line in  G09.83808, we re-observed it with ALMA Band~9 (Project \#2019.1.01427.S). The observations were taken on 2019 December 5 and 9, giving a total on-source time of 99.3~min. The array consisted of 41 twelve-meter antennas with baseline lengths of 15--312~m, providing sensitivity to structure on scales of 0.2--6.0~arcsec. The December 5 observations were of poor quality and were discarded. The December 9 data were taken in outstanding weather conditions, with a precipitable water vapour of 0.3~mm. 

The ALMA observations were reduced and imaged using {\sc CASA} v5.6.1 \citep{CASA2022}. We flagged antennas and frequency channels with elevated noise. The resulting FWHM synthesised beam size is 0.62'' $\times$ 0.41'' (natural weighting); the continuum imaging reaches a 1~$\sigma$ noise level of 0.53~mJy/beam.

\begin{figure*}
    \centering

    \includegraphics[width = 0.99\textwidth] {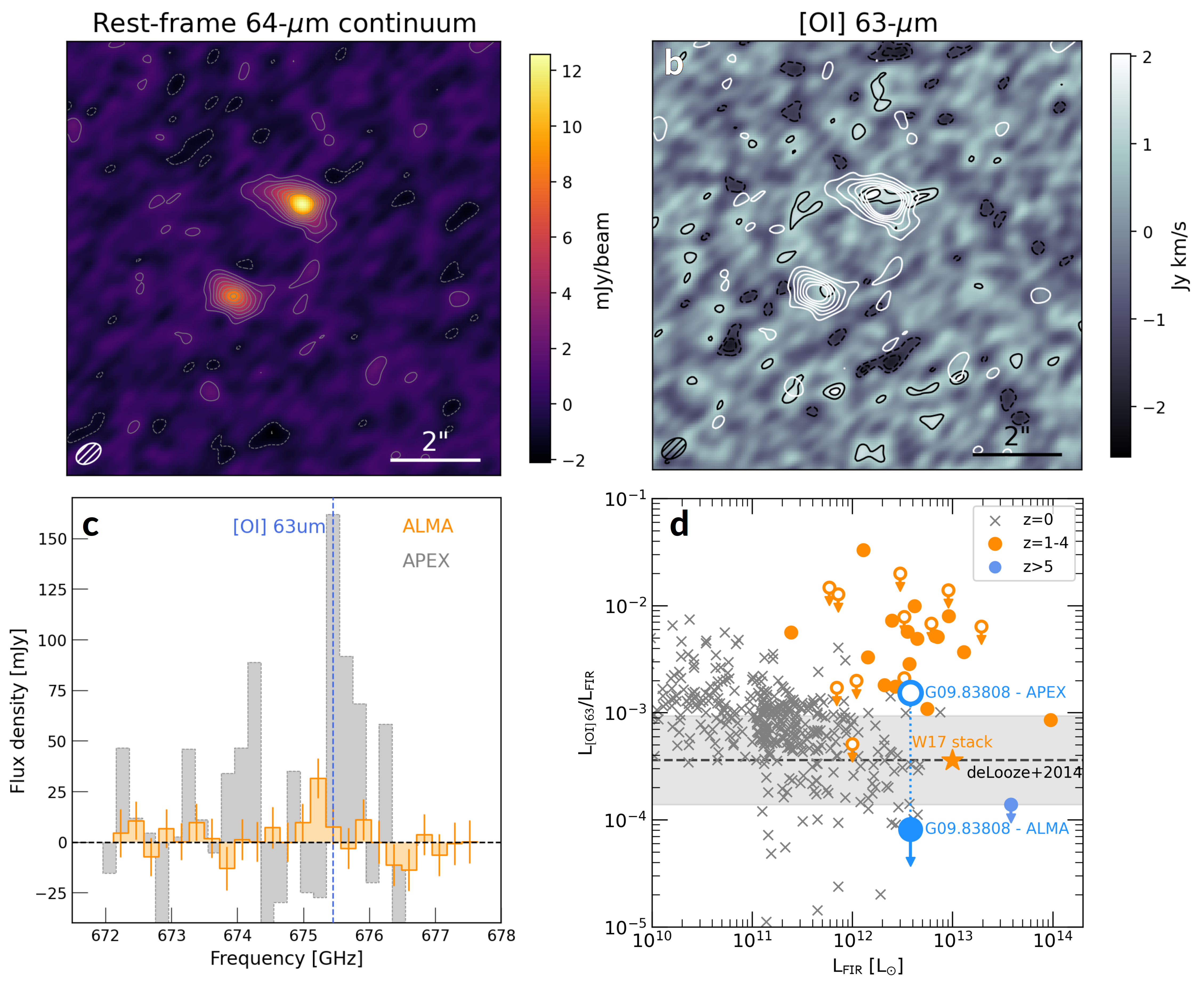}
    \caption{a): \textsc{Clean}ed image of the 64-$\mu$m rest-frame continuum, using natural weighting. The contours are drawn at ($\pm$2, $\pm$4, $\pm$6, ...)$\sigma$. 
    b) Dirty moment-0 image at the frequency of the [\ion{O}{1}]$_{63}$; black contours drawn at ($\pm$2, $\pm$3, $\pm$4, ...)$\sigma$. White contours indicate the Band~9 continuum.
    c) ALMA spectrum (orange) extracted from the two lensed images. We do not find any significant line emission, indicating that the claimed detection in \citet{Rybak2020a} (APEX, grey) is spurious.
    d) [\ion{O}{1}]$_{63}$/$L_\mathrm{FIR}$ ratio in G09.83808 (blue circles) compared to literature data and the \citet{deLooze2014} empirical relation.}
    \label{fig:1}
\end{figure*}

Figure~\ref{fig:1} (left) shows the \textsc{Clean}ed rest-frame 64-$\mu$m continuum image: G09.83808 is resolved into two lensed images with a peak S/N of $\approx$25. The total continuum flux is $S_\mathrm{64\mu m}=33.8\pm1.6$ mJy, consistent with the archival photometry \citep{Tadaki2022}.

We then subtracted the continuum using a first-order polynomial fit to the line-free SPWs. Fig.~\ref{fig:1}b shows a dirty-image map at the expected position of the [\ion{O}{1}]$_{63}$ line collapsed over 260 km/s, twice the line FWHM reported in \citetalias{Rybak2020a}; Fig~\ref{fig:1}c shows the spectrum extracted from the two lensed images. We do not see any significant [\ion{O}{1}]$_{63}$ emission: neither does $uv$-plane analysis reveal any excess emission at the expected frequency of the [\ion{O}{1}]$_\mathrm{63}$ line.  
The non-detection is robust with respect to calibration and imaging procedures: the tentative ($\approx$3$\sigma$) emission seen in Fig~\ref{fig:1}b disappears for different bandwidths and $uv$-tapers. 

\begin{figure*}[t]
    \centering

    \includegraphics[width = 0.49\textwidth] {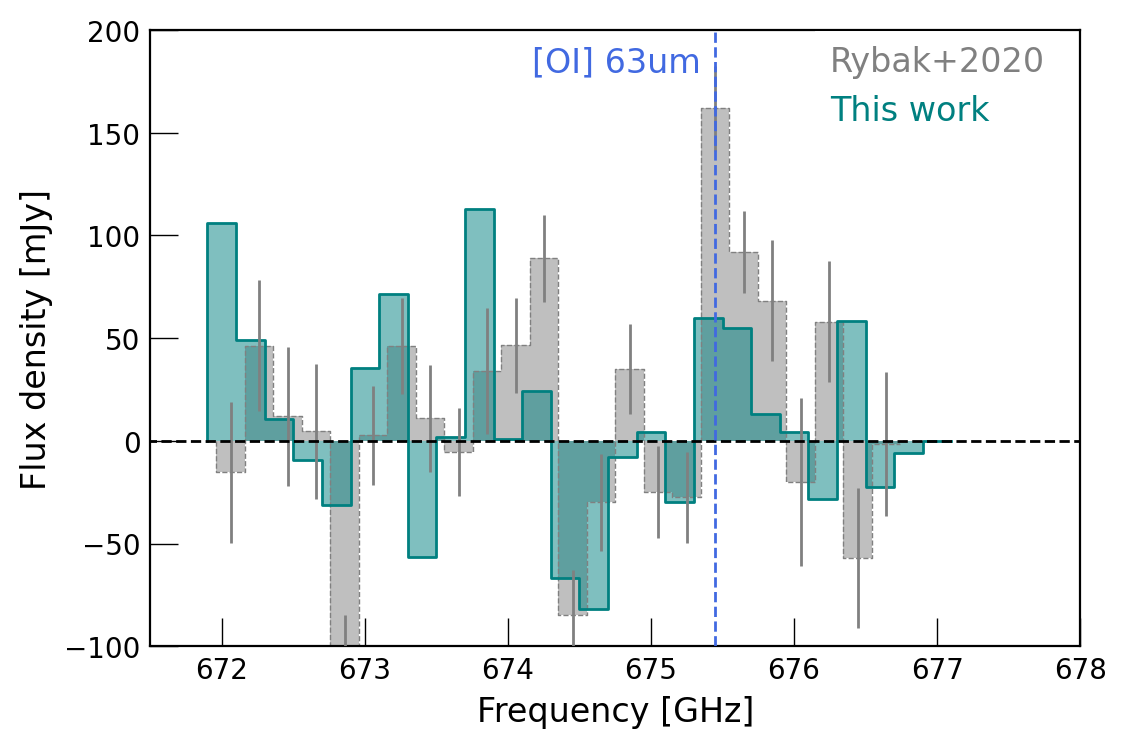}
    \caption{{Comparison of the APEX SEPIA660 spectra from \citet[grey]{Rybak2020a} based on the 2019 October 28 and November 11 observations, and a new reduction of 2019 November 11 and December 11 observations (teal). Our new reduction does not confirm the claimed detection of the [\ion{O}{1}]$_{63}$ line.}}
    \label{fig:2}
\end{figure*}

Finally, we re-analyse the original APEX data from \citetalias{Rybak2020a}(2019 October 28 and November 6). After the publication of \citetalias{Rybak2020a}, additional data were taken on 2019 December 11. Compared to the other two observing blocks, the data from October 28 are affected by the higher pwv. Reducing the data from November 6 and December 11 only, does not yield a significant detection despite a lower rms
%\textit
{(Fig.~\ref{fig:2})}. 

\section{Results}

We set a 3$\sigma$ upper limit on the [\ion{O}{1}]$_{63}$ flux $\leq$6.16~Jy km s$^{-1}$, corresponding to an apparent luminosity of $L_\mathrm{[OI]63}^\mathrm{sky}\leq2.9\times10^9$~$L_\odot$ -- almost 20$\times$ lower than the value reported in \citetalias[$(54\pm12)\times10^9$~$L_\odot$]{Rybak2020a}.  After correcting for the lensing magnification ($\mu=9.3$, \citealt{Zavala2018}, we obtain a source-plane upper limit of $L_\mathrm{[OI]63}\leq3.1\times10^8$~$L_\odot$. As the [\ion{O}{3}]$_\mathrm{88}$ and CO(2--1) lines have extent similar to the dust continuum \citep{Tadaki2022, Zavala2022}: $\approx$2~arcsec, significantly smaller than the maximum recoverable angular scale of 6~arcseconds of our ALMA data. It is thus unlikely that the [\ion{O}{1}]$)\mathrm{63}$ is so extended as to be resolved out with ALMA.

To put our non-detection in context, Fig.~\ref{fig:1}d shows the ratio between [\ion{O}{1}] and far-infrared luminosity $L_\mathrm{FIR}$ as a function of $L_\mathrm{FIR}$ for other high-redshift \citep{Coppin2012, Brisbin2015, Wardlow2017, Wagg2020, Litke2022} and low-redshift \citep{GraciaCarpio2011, Coppin2012,Diaz2017} observations. Our ALMA-derived upper limit for G09.83808 is significantly lower than previous high-redshift detections but comparable to the brightest $z\sim0$ ultraluminous infrared galaxies (ULIRGs) and the upper limit for the $z=5.7$ dusty galaxy SPT~0346-52 \citep{Litke2022}. Given the small magnification gradients across the source, the [\ion{O}{1}]/FIR ratio is not significantly affected by differential magnification. The faintness of the [\ion{O}{1}]$_\mathrm{63}$ emission from G09.83808 ($\geq$5$\times$ fainter than either [\ion{C}{2}]$_{158}$ or [\ion{O}{3}]$_\mathrm{88}$)
might be caused by large optical depth and self-absorption by the foreground cold gas, as seen in some $z\sim0$ ULIRGs (e.g., \citealt{Farrah2013, Diaz2017}). The spurious $\geq5\sigma$ APEX detection of the [\ion{O}{1}]$_\mathrm{63}$ highlights the challenges of single-dish spectroscopy in high-frequency sub-millimetre bands.

\begin{acknowledgments}

This paper makes use of the following ALMA data: ADS/JAO.ALMA\#2019.1.01427.S. ALMA is a partnership of ESO (representing its member states), NSF (USA) and NINS (Japan), together with NRC (Canada), MOST and ASIAA (Taiwan), and KASI (Republic of Korea), in cooperation with the Republic of Chile. The Joint ALMA Observatory is operated by ESO, AUI/NRAO and NAOJ.

%\textit
{This publication is based on data acquired with the Atacama Pathfinder Experiment (APEX) under programme ID 0104.B-0551. APEX is a collaboration between the Max-Planck-Institut f\"ur Radioastronomie, the European Southern Observatory, and the Onsala Space Observatory.}

The authors acknowledge assistance from Allegro, the European ALMA Regional Center node in the Netherlands.

M. R. is supported by the NWO Veni project "\textit{Under the lens}" (VI.Veni.202.225) and the Delft Space Institute Seed Grant "\textit{Hundred colours of galaxies}".
%\textit
{J. L. W. acknowledges support from an STFC Ernest Rutherford Fellowship (ST/P004784/2).}
      
\end{acknowledgments}

\vspace{5mm}
\facilities{ALMA, APEX}

\bibliography{references}{}
\bibliographystyle{aasjournal}

\end{document}